\documentclass[12pt]{article}
\usepackage{amsfonts,amssymb,latexsym,amscd}
\title{\textbf{The Interacting and Non-constant Cosmological Constant}}
\date{}
\author{ Murli Manohar Verma \footnote{sunilmmv@yahoo.com, mverma0@ictp.it}\\
{\small Department  of  Physics}\\
{\small Lucknow  University,  Lucknow  226 007, India}\\
{\small and}\\
{\small The Abdus  Salam  International  Centre  for  Theoretical Physics}\\
{\small Trieste,  Italy}\\
}

\begin{document}
\maketitle


 \abstract

 {We propose a time-varying cosmological constant with a fixed equation of state,  which evolves mainly   through its interaction   with the background   during most of the long history of the universe. However, such interaction does not exist in the very early and the late-time universe and produces the acceleration during these eras  when it becomes very nearly a constant. It is found that after the initial inflationary phase, the cosmological constant, that we call as lambda parameter, rolls down from a large constant value to another but very small  constant value and further  dominates the present epoch showing up in form of the dark energy driving the acceleration.}


\vskip.1in
\noindent
 PACS:

\pagebreak

\section{Introduction}The question to determine the nature and evolution of the dark energy in the universe has been the focus of the perplexity in physics over the past decade since its inception through the high redshift supernovae \cite{a1,b1} and the Cosmic Microwave Background (CMBR) observations \cite{c1}. To explain these  observations in precision cosmology, many  possibilities are being investigated into.  The  search for various alternative approaches  for dark energy mainly includes the minimally coupled \cite{c2,d1,e1} or the interacting scalar field quintessence models  \cite{f1,f2}  and the cosmological constant \cite{f3}. Some authors  have  proposed a single scalar field that produces the early and the late-time inflation \cite{g1}, or the thawing and freezing models \cite{g2}. On the other hand, the cosmological constant stands as a viable alternative in $\lambda$CDM model \cite{g4,g5},  with the present observations giving an equation of state (EOS) converging towards $w\sim -1$ ($w=-0.969\pm0.061$(stat)$\pm 0.065$(sys)\cite{h2}). However, this raises another equally mysterious  question \emph{i.e.,}  the cosmological constant  problem, whose solution is still far from  attained \cite{g4,g5,h2,h4,h5,h6,h7}.

  In this paper, we study the cosmological constant from a different approach. In our opinion, the `correct' form of the action must be explored from the dynamics of the field instead of fixing it at the beginning. Intuitively, it is hard to believe that such a significant and all-pervasive component, in form of the cosmological constant, and perhaps dominant as vacuum energy \cite{g5,h4,h5,h6,h7},  both at small and the cosmic  scales,  must  lie dormant throughout its long history without ever interacting with the matter and still drive  the dynamics of the universe. Having this motivation,  we study  the dynamics of an interacting cosmological constant (we will call it a  `parameter'  when interacting) which drives the inflation at different epochs and relaxes in-between. Simple considerations lead to a common behaviour of  the interacting cosmological parameter $\lambda$ with a fixed EOS and the quintessence with an  evolving  EOS , which is, however, easily distinguished from  the non-interacting cosmological constant.

 With this view, we study an alternative scenario of a three-phase cosmological evolution. In the earliest  phase I and the latest phase III,  we have a universe with a non-interacting  cosmological constant $\lambda$ along with the radiation or matter,  each  evolving separately. However, in the intermediate phase II (which we call as  Q phase),  it interacts with the background matter. Thus in phases I and III,  the divergence of stress-energy tensor vanishes for each component \emph{i.e.}  $T^{ik}_{;k}=0$ ,  with the matter energy density varying with the scale factor $a(t)$ as $\rho_{n}\propto a^{-3(1+w_{n})}$,  $w_n $ being its EOS,  and  the energy density in form of the cosmological constant $\rho_{\lambda}$ = constant  while in Q phase,  $(T^{ik}_{(n)}+T^{ik}_{(\lambda)})_{;k}=0$  and the scale dependence of energy densities is influenced by the interaction. Consequently, as we will see in Section 2,  $\rho_{\lambda}$ falls faster than $\rho_{n}$ before it enters the next (present) phase III and subsequently dominates over matter $(\rho_{\lambda}/\rho_{n}=r\geq1)$.

 A part of  our motivation for discussing  the scale-dependent $\lambda$  stems from the simple curvature consideration as follows. Writing $\lambda$ term with curvature in the  Friedmann  equations (with units using $c=1$ )
\begin{eqnarray}(1-2q) H^2+\frac{k-\lambda a^2/3}{a^2}&=& 8\pi G T_{\mu}^\mu \label{n1}\end{eqnarray}
\begin{eqnarray}H^2+\frac{k-\lambda a^2/3}{a^2}&=&
\frac{8\pi G}{3}T_{0}^0\label{n2}\end{eqnarray}(where $q(t)=-\ddot{a}/aH^{2}(t)$ is the deceleration parameter, $\mu =1,2,3$ and other terms have their usual meaning)
 we have \begin{eqnarray}k_{\emph{eff}}=k- \lambda a^2/3.\label{n3}\end{eqnarray}
 If we put $k_{\emph{eff}}\sim 0$ as suggested by the CMBR data by sufficient margin of error (systematic or statistical) \cite{c1} then  $k\sim 0=$constant 
indicates $a^{-2}$ dependence of $\lambda$. In the matter dominated (MD) universe,  it gives $\lambda \propto H^{4/3}$ and in the radiation dominated (RD) universe,  $\lambda \propto H$. A few other suggestions indicate the ansatz $\lambda \propto H^{4}$ in the quasi-de Sitter phase of early inflation prior to the RD era \cite{i1}. The late-time vacuum induced by the quantum condensates giving a  $\lambda \propto H$ has also been proposed \cite{j1}. It is understood that the vacuum effects must be clearly distinguished from the  ``$\lambda$ only '' effects,  which can indeed be now  so small as $L_{\lambda}/L_{P}\sim 10^{61}$ (with $L_{\lambda}$ and $L_{P}$ as the $\lambda$-  and the Planck length scales respectively) depending on the present observed values of Hubble parameter $H_{0}$ and its  closure  parameter $\Omega_{\lambda 0}$. Therefore, in our concerns about cutting down the size of $\lambda L_{P}^2$  to a value  $\sim 10^{-123}$,  it would be quite appropriate to consider its scale dependence which is,  in fact,   not ruled out in cosmology (unless, of course, it is taken as a constant $\emph{per se}$).

However,  (\ref{n3}) is only suggestive in nature and it is also not clear how a global geometry can evolve despite our knowledge of the local curvature obtained  within  the limited scope of observations (even with the acquisition of high redshift data) \cite{j2} in the absence of our faith in a working  cosmological principle \cite{k1,l1}.

In Section 2,  we discuss the dynamics of interacting cosmological parameter $\lambda$ in Q phase.  This parameter (which is  no longer constant, except for its  EOS) plays a role like the interacting quintessence but with fixed EOS equal to $-1$, and relaxes towards a small value because of the interaction with the background matter.

The next Section 3 includes a discussion on the  dignostics suggested by some authors primarily to distinguish the cosmological constant from quintessence with  varying $w(z)$ \cite{l2,m2}.  However, we find that the same diagnostics (which may include the other functional forms  of the third derivatives of the scale factor) are useful  in a $\lambda$CDM universe  to distinguish between interacting and non-interacting cosmological constant too. We attempt to constrain the evolution of the $\lambda$ that indicates a solution to the observed smallness of the  cosmological constant, and that of the ratio of energy densities $r$ in the main components, $\lambda$ and matter. Finally, in Section 4, we give a summary of our conclusions.

\section{Initial conditions and the evolution of cosmological constant}

A set of initial conditions are possible  in phase I,  depending on which, $r$ rises with the scale factor of the universe  (\emph{e.g.} from $r<1$ to $ r>1$ when it evolves from an RD era into a $\lambda$ dominated ($\lambda$D) era) or  up from $ r>1$ if the universe is $\lambda$D  initially,  followed by an  RD era. This growth stops soon at an epoch  when it  enters Q phase to be discussed below. This  set offers a wide class of possible evolutionary tracks of the universe.  For example, in phase I, we have two possibilities (i) RD  $\rightarrow \lambda$D transition or (ii) $\lambda$D alone. However, an  evolution initiated with  a very large  cosmological constant will provide with a too short RD era in the following Q phase and hence  little time for the nucleosynthesis to complete, against the observations of  the  abundance of light nuclei. On the other hand,  its too small value will make it incompatible with the scale of the standard inflation near GUT scale $\sim 10^{16}$ GeV.

Following the normal initial inflation of the phase I,  if we include the possibility of an interacting cosmological parameter during  an intermediate Q phase of the universe, then the conservation of energy gives
\begin{eqnarray}\dot{\rho}_{n}+3H \rho_{n} (1+w_{n})&=& Q\label{n4}\end{eqnarray}
\begin{eqnarray}\dot{\rho}_{\lambda} &=& -Q\label{n5}\end{eqnarray}
where  $w_n=0$ for matter, 1/3 for radiation,  and $ Q $  determines the strength of interaction mainly between $\lambda$ and the  dominant background, radiation or matter. While a possibility of changing `constants' like c or G will  also result in the energy density variations, we do not consider it here.

In phases I and III,  $ Q=0$,   and  $ \rho_{\lambda}= \lambda / 8\pi G $ = constant,  as would be the normal behaviour of $\lambda$ with its EOS  $w_{\lambda} =-1$. However, in  Q phase,  $Q\neq 0$  and there exist three possibilities. If we retain the same EOS of the cosmological constant  throughout then the above interaction manifests itself in the time-variations of the energy density. Alternatively, it can be seen that in the opposite case,  $ w_{\lambda}=-\alpha Q/H -1 $  where $\alpha$ is a constant with a large value depending on $G$ and $\lambda$.  It puts  a constraint on $ w_{\lambda}$ through the interaction  and  makes  it much different from its standard  value of  $-1$. In addition to it, it does not solve the cosmological constant problem since $\lambda$ always remains a constant during the evolution. The third possibility, of course, may be the variation of both $\lambda$ and $ w_{\lambda}$ in such a way that as to provide with a constant or time-dependent  interaction and this must be addressed separately.

In Q phase, with (\ref{n4}) in the background,  the dynamics of an interacting $\lambda$,  now behaving  like the spatially homogeneous interacting scalar field $\phi$  but with EOS  equal to $-1$, is given by
\begin{eqnarray}\ddot{\phi}+3H \dot{\phi} +V'(\phi) &=& - \frac{Q}{\dot{\phi}}\label{n6}\end{eqnarray}
\begin{eqnarray}Q &=& \frac{\rho_{n2}}{f_{2}}\left(\frac{a_{2}}{a}\right)^{3(1+w_n)}\dot{f}(\phi)\label{n7}\end{eqnarray}
where the subscript $2$  in (\ref{n7})  denotes  the corresponding values at the end of Q phase. Thus $f_{2}=1$ as also in the presently ongoing  phase III.

  Solutions to (\ref{n4}) and  (\ref{n5}) yield  $\rho_{n}\propto a^{Q/ H \rho_{n}} a^{-3(1+w_{n})}$ that makes the evolution of  matter energy density fall in a concave manner much slower than what we would expect in  a non-interacting case. This  is because of the coupling function $f(\lambda/M_{P}) \equiv a^{Q/H\rho_n}$ between $\lambda$ and the background matter whose particles have their masses influenced by dark energy \cite{l3}. It can also be seen that in Q phase  $\rho_{\lambda}\propto  a^{-Q/ H \rho_{\lambda}}$.  Note the role played by the interaction term Q in the evolution of the corresponding component (radiation, matter or $\lambda$). We find that $\rho_{\lambda}$ falls at a much faster rate than its  ``loitering''  matter counterpart.

  The time-dependence in the ratio of energy densities  $r=\rho_\lambda/\rho_n$ enters through the coupling function $f$.  From the solutions of (\ref{n4}) and  (\ref{n5}),  we can get its functional dependence as $rf^{1+1/r}\propto a^4 $ in RD (and $\propto a^3$ in MD) universe. In terms of $r$  we find $\rho_\lambda \propto f^{-1/r}$.

  Since for each background component  $\rho_n \propto t^{-2}$,  the time evolution of $f$ in (\ref{n7}) is given by $f\propto t^3$ assuming $Q$ as constant.

In terms of the present energy density $\rho_{n0}$,  (\ref{n7}) can be expressed as
\begin{eqnarray}Q &=& \rho_{n0}(1+z)^{3(1+w_{n})}\dot{f}(\phi)\label{n8}\end{eqnarray}
where we have, as suggested by the present CMBR,  BAO and  other observational constraints on baryonic and dark matter,  $\Omega_{n0}\sim 1-\Omega_{\lambda0}$ giving  $\rho_{n0}\sim H_{0}^2 M_{P}^2$,  with $\Omega_{\lambda0}\sim 0.7$ \cite{c1,h2}.

Thus for  $\lambda$ with  $w_{\lambda}=-1$,   $\dot{\phi}\sim0$, we have the large interaction term $(-Q/\dot{\phi})$ on the right-hand side of (\ref{n6}).  Since $V'(\phi)=-Q/\dot{\phi}$,  this means that even a moderate  value of $Q$ is enough in this case for the fast descent of the potential and can alleviate the cosmological constant to a very small observed value following
\begin{eqnarray}\frac{V'(\phi)}{V}= - \frac{Q}{3\dot{\phi}H^2 M_{P}^2 \Omega_{\lambda}}\label{n9}\end{eqnarray}
where $\Omega_\lambda$ is  its closure parameter and $M_P = (8\pi G)^{-1/2}$ is the reduced Planck mass.
  From (\ref{n8}) and (\ref{n9}) it is clear that $f'/f$ puts a natural control on the evolution of potential $V(\phi)$

The interacting $\lambda$ is equivalent to non-interacting components, each with a varying EOS parameter (resulting in the corresponding pressure shift), which appear from (\ref{n4}) and (\ref{n5}) as
 \begin{eqnarray}w_{n\emph{eff}}=  w_{n}-\frac{Q}{3H \rho_{n}}. \label{n10}\end{eqnarray}
 \begin{eqnarray}w_{\lambda\emph{eff}}=-1+\frac {1}{3r}\left(\frac{\ln f}{\ln a}\right). \label{n11}\end{eqnarray}

  In presence of  interaction, the  effective EOS is given as
 \begin{eqnarray}w_{\emph{eff}}=  w_{n}\Omega_{n}+w_{\lambda}\Omega_{\lambda} \label{n12}\end{eqnarray}
   which becomes $w_{\emph{eff}}= -\Omega_\lambda$ in case of dust.

\section{The diagnostics for  $\lambda$-matter  interaction}

   In a spatially flat universe,  the deceleration parameter in (\ref{n1}) can be used to distinguish $\lambda$  from quintessence with a varying EOS (at present $w_{\phi0}\sim \frac{1}{2}(2q_0-1)$ with $\Omega_{\phi0}\sim 0.7$, and if dark energy is indeed a cosmological constant,  we have $q_0\sim - 0.5$).  However,  $q(t)$  is not sensitive to the interaction strength $Q$,  where the third derivatives of the scale factor,  ( with  $\alpha=\dot{a} = aH$ ),   are effective in form of the following statefinders \cite{l2}
  \begin{eqnarray}u=\frac{\ddot{\alpha}}{aH^3}\label{n13}\end{eqnarray}
 \begin{eqnarray}s=\frac{u-1}{3(q-1/2)}
 \label{n14}\end{eqnarray}
 which reduce, for $w_{\lambda}=-1$,  to
  \begin{eqnarray}u=1-\frac{9}{2}\left(\frac{Q}{3H\rho}\right) \label{n15}\end{eqnarray}
 \begin{eqnarray}s=\frac{Q}{3H\rho_\lambda} \label{n16}\end{eqnarray}
 where in (\ref{n15}) $\rho= \rho_n+ \rho_\lambda$ is the total energy density. Here,  $s$ is particularly sensitive to  any variations in the cosmological parameter $\lambda$.
 Thus in terms of $u-s$  space  for Q phase,  we get from (\ref{n4}), (\ref{n11}) and (\ref{n16}),  $f\propto a^{3rs}$ and $w_{\lambda \emph{eff}}=-1+s$.  In the non-interacting phases I and III of the $\lambda$CDM  universe,  we have $s =0$,  but in Q phase it varies with non-zero value. This effectively measures the ``loitering'' (mentioned in the previous Section)  of matter energy density much before the epoch of $\lambda$-matter equality and must provide with a lower background energy density prior to the end of Q phase, \emph{e.g.} for a given $\rho_{n2}$ in (\ref{n7}), than expected in the absence of interaction.  Therefore, it must play a significant role in  the subsequent
 structure formation ahead. Observationally,  this is also crucial for any model that generates dark matter through an interaction with dark energy \cite{m1}.

 To break the degeneracy in  the $u-s$ space, we use the requirement that $\lambda$ energy density must fall faster than the background. In the RD universe, this implies
 \begin{eqnarray}\frac{Q}{H\rho_\lambda}>4 \label{n17}\end{eqnarray}with
 \begin{eqnarray}\rho_\lambda < \rho_{\lambda 1}\exp [{-4H(t-t_1)}]\label{n18}\end{eqnarray}
 giving $ u<1-6\Omega_\lambda$ and $s>4/3$,(subscript 1 in  (\ref{n18})  denotes the initial moment of Q phase)
 while the corresponding conditions in the MD universe, with the right-hand sides of (\ref{n17}) and (\ref{n18}) changing to 3 and $\rho_{\lambda 1}\exp[-3H(t-t_1)] $ respectively,  give $ u<1-\frac {9}{2}\Omega_\lambda$ and $s>1$. Specifically, with  $u=0$, it puts an upper bound  $\lambda< H^2/2$   for RD  ( and $<2H^2/3$ for MD) universe.  Since  (\ref{n17}) is a stronger condition,  it can be used to constrain the interaction strength $Q$ around the scale of initial inflation.

 Using (\ref{n18}),  we can also constrain the time evolution of the ratio of energy densities $r$ discussed earlier, as
 \begin{eqnarray}\dot{r}<2t\exp[-4Ht(1-2Ht)].\label{n19}\end{eqnarray}

 With a view to distinguish the constant EOS of the cosmological constant from that of  the quintessence, a  new promising  diagnostic $Om$  has been introduced in terms of the Hubble parameter $H(z)$ as a function of the redshift $z$ ($H_0$ being its present value)  as \cite{m2}
 \begin{eqnarray}Om(z)=\frac {(H(z)/H_0)^2-1}{(1+z)^3-1}. \label{n20}\end{eqnarray}
 In the intermediate Q phase of universe,  the $\lambda$-matter coupling naturally enters (\ref{n20}) which now  becomes
 \begin{eqnarray}Om(z)= \Omega_{n0}f + \frac {(1- \Omega_{n0})(1+z)^{Q /H \rho_\lambda}+\Omega_{n0}f-1}{(1+z)^3-1} \label{n21}\end{eqnarray}
 where $\Omega_{n0}$ is the present  closure  parameter of matter.

 Clearly, in the absence of interaction,  $Om(z)$ reduces to $\Omega_{n0}$ as expected,  which will rule out a Q phase,  whereas any deviation from this value in the past will indicate an interacting $\lambda$ in the $\lambda$CDM universe.

\section{Summary}We have attempted to use an unconventional approach (not via fixing the action \emph{ab initio}) with some basic motivation for an underlying interacting cosmological constant (parameter) during  an intermediate Q phase in the universe. The dividend of this interaction is an eventually  relaxed cosmological constant. We have discussed this possibility in a three-phase evolution of the universe. The phases I (earliest) and III (late-time) have all common features of the  usual cosmological constant generating acceleration, except those embedded onwards as a result of the interaction in Q phase. However, as discussed in Section 2,  during  the entire Q phase  the matter energy  ``loiters''  in a slow roll, while the $\lambda$  decays  swiftly to achieve a small value due to the mutual coupling. This not only relaxes $\lambda$ but also seems to suggest some mechanism for matter creation from this decaying $\lambda$. This, together with our conclusion  about  any loitering behaviour that gives a lower densities,  both in radiation and matter during and prior to Q phase,  must have significant observational consequences to be studied in future.

In Section 3, we have examined the diagnostics for the interacting $\lambda$ and constrained its (and of $r$)  evolution.  However, some more pointed functional forms including the third or higher derivatives may be studied  which break the degeneracy in case of the above-mentioned embeddings along the ride.

\vskip.2in
\noindent
\textbf{\large Acknowledgments }\\

\noindent
The author is thankful to Paolo Creminelli for useful discussions and to  The Abdus Salam International Centre for Theoretical  Physics, Trieste, Italy, for its facilities. This work was completed under the Federation Arrangement.

\end{document}